\documentclass[aps,prc,twocolumn,showpacs]{revtex4}
\usepackage{graphicx}
\begin{document}

\title{Scalar ground-state observables in the random phase approximation}
\author{Calvin W.~Johnson}
\affiliation{Department of Physics,
San Diego State University,
5500 Campanile Drive, San Diego CA 92182-1233}
\author{Ionel Stetcu}
\altaffiliation{On leave from National Institute for Physics and
Nuclear Engineering -- Horia Hulubei, Bucharest, Romania.}
\affiliation{
Department of Physics and Astronomy,
Louisiana State University,
Baton Rouge, LA 70803-4001
}

\begin{abstract}
We calculate the ground-state expectation value of scalar
observables in the matrix formulation of the random phase
approximation (RPA). Our expression, derived using the quasiboson 
approximation, is a straightforward generalization of
the RPA correlation energy.  We test the reliability of
our expression by comparing against full diagonalization in  $0\hbar\omega$
shell-model spaces.  In general the RPA values are an improvement
over mean-field (Hartree-Fock) results, but are not always consistent 
with shell-model results.  We also consider exact symmetries
broken in the mean-field state and whether or not they are restored
in RPA.
\end{abstract}

\maketitle

\section{introduction}

Mean-field theory is the starting point for most microscopic
models of many-body systems such as nuclei.  In fact, for
\textit{global} studies of nuclear properties, such as nuclear binding
energies, mean-field theory in the form of Hartree-Fock (HF) and 
related approximations are still the only viable microscopic approach \cite{HFmass};
otherwise one turns to semiclassical methods such as liquid drop \cite{MN95} and
Thomas-Fermi \cite{AP95}.

Unfortunately, mean-field theory
ignores particle-hole correlations and can break
fundamental symmetries such as rotational and translational
invariance.  The next logical step beyond the static mean-field is
known as the random phase approximation, or RPA, which can be
derived in the small amplitude limit of time-dependent
mean-field theory, and which implicitly
allows for small correlations in the ground state \cite{Rowe70,ring,BB}.

The main applications of RPA have been to transition strengths
and excitation spectra.  One can also calculate the correction to
the ground state energy due to RPA correlations, which was the
basis of a recent proposal to use mean-field  + RPA
to compute global binding energy systematics \cite{bertsch2000}.

There are other potential quantities of interest, namely the ground state
expectation value of observables other than the Hamiltonian.
An important application is the root-mean-square radius.  Any global fit of nuclear systematics
involves both binding energies and nuclear radii. Indeed,
most global fits to binding energies and radii with
nonrelativistic phenomenological interactions such as Skyrme require 
density-dependent forces, although we will not deal specifically with that issue
in this paper. Further challenges to mean-field calculations 
are isotopic shifts in the charge radius \cite{pa73},
and parity-violating electron scattering, which
is sensitive to the difference between the neutron and
proton rms radii \cite{pvrms}.

In section II of this paper we present an expression for the ground-state expectation value of a
general observable in RPA. This expression is derived using the quasiboson 
approximation, in exact analogy with the RPA correlation energy, and is 
a generalization of the RPA one-particle density derived by Rowe \cite{Rowe70,ro68}. 
Because the RPA ground state is approximated by a quasiboson vacuum, 
one can read off directly the expectation value without having to 
construct an explicit wavefunction.

Having an expression is not sufficient. Does it produce useful and
reliable results? After all, RPA is an approximation, and because it
violates the Pauli exclusion principle RPA is not even variational.
Despite this, rigorous tests of the accuracy of RPA calculations 
have been spotty. 
For example, although the RPA correlation energy has been in the
literature for decades, the only tests were in toy models \cite{toyrpa}. 
Recently we compared the RPA correlation
energy against exact shell-model results \cite{stetcu2002}. We
found that generally RPA gave very good agreement--albeit with some
significant failures which reduce the reliability of RPA binding
energies.

We continue this program by computing the ground-state expectation
value of observables both in exact shell-model calculations and in
RPA.  As discussed below, in this paper we limit
ourselves to scalar operators. Unfortunately in $0\hbar\omega$ harmonic 
oscillator model spaces $\langle R^2 \rangle $ is trivial, so we instead compute and compare
the expectation values of a one-body
operator: the number of
particles outside the $0d_{5/2}$ orbit; and several
two-body operators: $S^2$,
$L^2$, $Q^2$, and the pairing Hamiltonian $H_{\rm pair}=P^\dagger P$.

The results are found in Section III.  In general, RPA offers an improvement
over the mean-field (HF) expectation values.  This is not always the case,
however, and even when HF+RPA is closer to the exact shell-model results than
HF values, the improved agreement is sometimes still unsatisfactory.
(The situation is reminiscent of variational calculations, where one typically
gets much better agreement in energy than in the wavefunction. But again,
RPA is not a variational theory.)

In section IV we turn to another issue of relevance to RPA:
the restoration of symmetries, such as rotational invariance,
broken by the mean field.
We compute the expectation value of ${J}^2 $ and compare the 
exact shell-model result against the Hartree-Fock and RPA values. 
If RPA
``restores'' the broken symmetry then one expects that
either one regains the
correct ground state value of ${J}^2$ or gets
a much better value than the Hartree-Fock value.
We find that, although often $\langle J^2 \rangle$ is improved by RPA
corrections, one does not always regain a very good estimate of the
ground state angular momentum.  We interpret this to mean that RPA 
only approximately restores broken symmetries.

Nonetheless, we consider our results relevant.  
The RPA is not new, but we remind the reader that
for global approaches to nuclear structure the state of the art is
still mean-field theory; \textit{ab initio} methods can only be applied to
very light nuclei, and shell model diagonalization can only be
applied piecewise.   Consistent calculations using HF+RPA throughout
the nuclides is conceiveable, however, and this paper is
part of a larger program
to rigorously test the reliability of RPA as an approximation to much larger microscopic
calculations.

\section{corrections to a general operator in RPA}

Before we derive the expectation value of an operator in
RPA, we first begin with a brief pedagogical review of
the derivation of the RPA correlation energy, including
proper treatment of zero modes that correspond to broken
symmetries. The RPA expectation value of a general operator
then follows naturally.

There are many ways to derive RPA \cite{ring,BB}.  One approach
is to expand the energy about the Hartree-Fock miminum.
Let $| HF \rangle$ be the Hartree-Fock Slater determinant, and let
$| \Psi(\vec{Z}) \rangle = \exp( \sum_{mi} Z_{mi}^* \hat{c}^\dagger_m
\hat{c}_{i} )| HF \rangle$ be small perturbations from the Hartree-Fock
state. We follow the usual convention where $m,n$ denote single-particle states above 
the Fermi surface in the Hartree-Fock state, or ``particle'' states, 
while $i,j$ denote ``hole'' states below the Fermi surface. 
Then one can expand the energy surface $E(\vec{Z})$ in the vicinity of
the Hartree-Fock minimum in terms of the particle-hole amplitudes
$\{ Z_{mi} \}$ up to second order:
\begin{equation}
E(\vec{Z})=E_{HF} +
\vec{z}^\dagger \textbf{A} \vec{z} + \frac{1}{2} \vec{z}^\dagger \textbf{B} \vec{z}^*
+ \frac{1}{2} \vec{z}^T \textbf{B} \vec{z},
\label{EofZ}
\end{equation}
where
\begin{eqnarray}
A_{nj,mi} \equiv \left \langle HF  \left |
\left [ \hat{c}^\dagger_j \hat{c}_n , [\hat{H} , \hat{c}^\dagger_m
\hat{c}_i ] \right ]\right | HF  \right \rangle, \label{defA} \\
B_{nj,mi} \equiv \left \langle HF  \left |
\left [   [\hat{H} , \hat{c}^\dagger_n
\hat{c}_j ], \hat{c}^\dagger_m \hat{c}_i \right ]
\right | HF  \right \rangle.
\label{defB}
\end{eqnarray}
We restrict ourselves to real wavefunctions, so that \textbf{A}, \textbf{B} are 
also real. 
There is no linear term in (\ref{EofZ}) because
$\left \langle HF \left | \left [ \hat{H}, \hat{c}^\dagger_m \hat{c}_i \right ]
\right | HF \right \rangle =0$ by the definition of the Hartree-Fock state.
This quadratic surface can
be mapped to a harmonic oscillator, replacing the
$Z_{mi}$ by boson operators $b^\dagger_{mi}$
\begin{equation}
\hat{H}_B=E_{HF} +
\vec{b}^\dagger \textbf{A} \vec{b} + \frac{1}{2} \vec{b}^\dagger \textbf{B} \vec{b}^\dagger
+ \frac{1}{2} \vec{b} \textbf{B} \vec{b},
\label{HB1}
\end{equation}
which has analogous commutation relations, e.g.,
$A_{nj,mi} = \left ( 0  \left |
\left [ \hat{b}_{nj} , [\hat{H}_B , \hat{b}^\dagger_{mi}
 ] \right ]\right | 0  \right )$, etc.
The boson Hamiltonian can be written in matrix form, which
induces an additional constant term:
\begin{equation}
H_B=E_{HF}-\frac{1}{2}\textrm{Tr} A + \frac{1}{2} (\vec{b}^\dagger
\:\: \vec{b})\left(\begin{array}{cc}
\mathbf{A} & \mathbf{B}\\
\mathbf{B} & \mathbf{A}
\end{array}\right)\left(\begin{array}{c}
\vec{b} \\
\vec{b}^\dagger
\end{array}\right).
\label{HB}
\end{equation}

We want to put (\ref{HB}) into diagonal form, the first step of
which is to solve the well-known RPA matrix equation:
\begin{equation}
\left(
\begin{array}{cc}
\mathbf{A} & \mathbf{B} \\
-\mathbf{B} & -\mathbf{A}
\end{array}\right)
\left(
\begin{array}{c}
\vec{X}\\
\vec{Y}
\end{array}\right)=
\Omega
\left(
\begin{array}{c}
\vec{X}\\
\vec{Y}
\end{array}\right).
\label{RPAmatrix}
\end{equation}
The solutions have normalization $\vec{X}^2 - \vec{Y}^2 = 1$.

Before going on one must treat carefully the case
of zero modes, $\Omega_\nu = 0$.
Zero modes arise in RPA when an exact symmetry is
broken. Suppose the HF state is deformed. The HF
energy will not change as the orientation is
rotated; this invariance is reflected in $E(\vec{Z})$ and
resurfaces as RPA zero modes. (This is also a good
check of one's RPA codes.  One expects three zero-frequency 
modes for triaxially deformed states, and two zero 
modes for an axisymmetric deformed state: rotation
about an axis of symmetry takes one to the same, 
indistiguishable state.) One cannot define
properly normalized $\vec{X},\vec{Y}$ for zero modes.
Instead one introduces  
collective coordinates $\vec{Q}_\nu$ and conjugate momenta 
$\vec{P}_\nu$\cite{ring,Rowe70,weneser}, which satisfy 
\begin{eqnarray}
\textbf{A} \vec{P}_\nu - \textbf{B} \vec{P}^*_\nu = iM_\nu \Omega^2_\nu \vec{Q}_\nu, 
\nonumber \\
\textbf{A} \vec{Q}_\nu - \textbf{B} \vec{Q}^*_\nu = - \frac{i}{M_\nu}
\vec{P}_\nu.
\label{RPAqp}
\end{eqnarray}
Here $M_\nu$ is a constant, interpretable as mass or moment of inertia 
but whose value depends on the normalization of $P, Q$, which 
are only constrained by  
\begin{equation}
\vec{Q}^*_\lambda \cdot \vec{P}_\nu - \vec{Q}_\lambda \cdot \vec{P}^*_\nu
= i \delta_{\lambda \nu}.
\label{qpnorm}
\end{equation}
Now one obtains a generalized Bogoliubov transformation 
of the boson operators $\hat{b}$, $\hat{b}^\dagger$, not only 
to quasibosons $\hat{\beta}$, $\hat{\beta}^\dagger$, but also 
to the zero-mode collective coordinate operator $\hat{\cal Q}_\nu$ and
conjugate momentum  operator $\hat{\cal P}_\nu$:
\begin{eqnarray}
\hat{b}_{mi} = \sum_{\lambda, \Omega_\lambda > 0}
\left( X_{mi,\lambda} \hat{\beta}_\lambda + Y_{mi, \lambda}
\hat{\beta}^\dagger_\lambda \right )
\label{boson_bogolyubov} \\
- i  \sum_{\nu, \Omega_\nu = 0}
\left (
P_{mi,\nu} \hat{\cal Q}_\nu - Q_{mi,\nu} \hat{\cal P}_\nu \right ).
\nonumber
\end{eqnarray}

Application of (\ref{boson_bogolyubov}) and its Hermitian 
conjugate to (\ref{HB}) puts it in diagonal
form:
\begin{equation}
H_B=E_{RPA}+\sum_\lambda \Omega_\lambda \beta_{\lambda}^\dagger\beta_\lambda
+ \sum_{\nu, \Omega_\nu = 0} \frac{\hat{\cal P}^2_\nu}{2 M_\nu}.
\end{equation}
The ground state is the quasiboson vacuum, and one can easily 
read off the RPA ground state energy, which is just the zero-point 
energy: 
\begin{equation}
E_{RPA}=E_{HF}-\frac{1}{2}\textrm{Tr}(A)+ \frac{1}{2} \textrm{Tr} (\Omega).
\label{ERPA}
\end{equation}

One can expand
\begin{equation}
\textbf{A} = \textbf{X} {\Omega} \textbf{X}^\dagger
+ \textbf{P}\textbf{M}^{-1} \textbf{P}^\dagger,
\end{equation}
where the second term is restricted to zero modes, and then obtain an
expression equivalent to (\ref{ERPA}),
\begin{equation}
E_{RPA} = -\sum_\lambda \Omega_\lambda | \vec{Y}_{\lambda}|^2
- \frac{1}{2} \sum_{\nu, \Omega_\nu=0} \frac{1}{M_\nu}
|\vec{P}_{\nu}|^2,
\label{ERPA0}
\end{equation}
an expression which explicitly segregates out the contribution from
zero modes.

The above derivation of the RPA correlation energy
$-\frac{1}{2}\textrm{Tr}(A)+ \frac{1}{2} \textrm{Tr} (\Omega)$ is well-known.
We derive the RPA expectation value of any generator operator $\hat{\cal O}$
in exactly analogous fashion.
Define $\tilde{A}$, $\tilde{B}$ as in  Eqs. (\ref{defA}) and
(\ref{defB}), respectively, by replacing the Hamiltonian by the observable
${\cal O}$, that is,
\begin{eqnarray}
\tilde{A}_{nj,mi}({\cal O}) \equiv \left \langle HF  \left |
\left [ \hat{c}^\dagger_j \hat{c}_n , [\hat{\cal O} , \hat{c}^\dagger_m
\hat{c}_i ] \right ]\right | HF  \right \rangle,  \\
\tilde{B}_{nj,mi}({\cal O}) \equiv \left \langle HF  \left |
\left [   [\hat{\cal O} , \hat{c}^\dagger_n
\hat{c}_j ], \hat{c}^\dagger_m \hat{c}_i \right ]
\right | HF  \right \rangle.
\end{eqnarray}
and define $o_{mi}=\langle HF|[c_m^\dagger
c_i,{\cal O}]|HF\rangle$, which does not in general vanish. (Because 
we restrict the Hartree-Fock state to real Slater determinants, 
$\tilde{A}, \tilde{B}$ and $o$ are real.) 
Please note that in computing $\tilde{A}$, $\tilde{B}$, and $o$ one
still uses the Hartree-Fock state from minimizing the Hamiltonian $\hat{H}$,
\textit{not} from minimizing $\hat{\cal O}$; we are finding the
expectation value $\langle \hat{\cal O} \rangle $ in the vicinity of the
minimum of $\langle \hat{H} \rangle$.

The form of the operator in the harmonic oscillator bosons is
\begin{eqnarray}
\hat{\cal O}_B = {\cal O}_{HF} - {1 \over 2} {\rm Tr \,} \tilde{\bf A}
+( \begin{array}{cc}
 \vec{o}, &  \vec{o}
\end{array} ) \cdot
\left ( \begin{array}{c}
 \vec{b} \\
 \vec{b}^\dagger
\end{array} \right ) 
\nonumber \\
+{1\over 2}
( \begin{array}{cc}
 \vec{b}^\dagger , &  \vec{b}
\end{array} ) \cdot
\left(
\begin{array}{cc}
\tilde{\bf A}  & \tilde{\bf B} \\
\tilde{\bf B}   & \tilde{\bf A}
\end{array}
\right ) \cdot
\left ( \begin{array}{c}
 \vec{b} \\
 \vec{b}^\dagger
\end{array} \right ),
\label{OB}
\end{eqnarray}

By transforming to the collective quasibosons (\ref{boson_bogolyubov}),
we again can trivially read off the quasiboson vacuum expectation 
value, which is the RPA
ground-state expectation value, without explicit construction 
of a wavefunction:
\begin{equation}
{\cal O}_{RPA} =  {\cal O}_{HF} - {1 \over 2} \textrm{Tr} \,\tilde{\bf A}
+{1\over 2} \textrm{Tr} \, {\bf \Theta},
\label{ORPA}
\end{equation}
where
\begin{equation}
{\bf \Theta} = {\bf X}^T\tilde{\bf A}{\bf X} + {\bf Y}^T
\tilde{\bf A}{\bf Y}
+ {\bf X}^T\tilde{\bf B}{\bf Y}+{\bf Y}^T\tilde{\bf B}{\bf X}.
\end{equation}
Substitution of Eq. (\ref{RPAmatrix}) with $A$, $B$ derived from
the Hamiltonian immediately regains the RPA binding energy (\ref{ERPA}).
It is important to emphasize again that the $X$, $Y$ used here are those calculated from
Eq.~(\ref{RPAmatrix}) using the \textit{original} $A$, $B$ matrices
(from the Hamiltonian); one does \textit{not} compute $X$, $Y$ using $\tilde{A}$, $\tilde{B}$.

As before we can rewrite (\ref{ORPA}) into an expression with explicit segregation of the
zero modes:
\begin{widetext}
\begin{eqnarray}
{\cal O}_{RPA} =  {\cal O}_{HF} & &+\sum_{\lambda (\Omega_\lambda > 0) }
\sum_{mi,nj} (\tilde{A}_{mi,nj} Y_{mi,\lambda} Y_{nj,\lambda}
+ \tilde{B}_{mi,nj} Y_{mi,\lambda} X_{nj,\lambda} ) \nonumber  \\
& &-{1 \over 2} i \sum_{\mu (\Omega_\mu = 0) }
\sum_{mi,nj} \tilde{A}_{mi,nj}
(P^*_{mi,\mu} Q_{nj,\mu} - Q^*_{mi,\mu} P_{nj,\mu} ).
\label{ORPA0}
\end{eqnarray}
\end{widetext}
We have confirmed numerically that (\ref{ORPA0}) yields the same values as 
(\ref{ORPA}).

As we will see in the next section, Eq. (\ref{ORPA}) reduces to that 
derived by Rowe for one-particle densities in spherical nuclei 
\cite{Rowe70,ro68}. 

If $\hat{\cal O}$ is a scalar, so that it is invariant under rotation, 
translation, etc., then the above is sufficient.  Nonscalar observables 
require a little more thought. In particular, consider 
observables that are angular momentum tensors, $\hat{\cal O}_{\kappa, \mu}$, with 
a nonzero rank $\kappa$. Here 
$\mu$ is the magnetic quantum number.  
If one applies the generalized Bogoliubov transformation (\ref{boson_bogolyubov}) 
then the linear terms in $o_{mi}$ do not vanish. The linear terms 
proportional to the quasiboson operators $\hat{\beta}$, $\hat{\beta}^\dagger$ 
are transitions to excited states; but, for nonscalars, there will be 
terms also proportional to $\hat{\cal Q}_\nu$ and to $\hat{\cal P}_\nu$. 
Furthermore, there will be quadratic terms, such as $\hat{\beta}^\dagger \hat{\beta}$, 
$\hat{\beta} \hat{\beta}$, etc., which manifestly do not contribute to the ground 
state, but also terms quadratic in $\hat{\cal Q}$, $\hat{\cal P}$.  
These terms arise from rotations (and in the more general case, translations, 
etc.): because they correspond to zero-frequency modes, they connect only to 
the ground state, but in a different orientation.  
If $\hat{\cal O}_{\kappa, \mu}$ is a spherical tensor, we know 
how it transforms under a rotation, and the linear and quadratic terms 
in $\hat{\cal Q}$, etc., are simply the linear and quadratic terms for 
small rotation angles. Nonetheless, the issue of nonscalar observables 
is not trivial, and we leave it to future work.

To summarize a first result of the present work, we have extended the
usual expansion of the Hamiltonian operator in the neighborhood of the
HF solution to any operator, with the purpose to incorporate RPA correlations
in computation of expectation values of relevant observables.
In the next section we test the reliability of this formula in a
non-trivial model, that is, the shell-model restricted space using realistic
Hamiltonians.

\section{RPA vs. shell-model for scalar observables}

In this section we test the RPA expectation value found in Eqs.~(\ref{ORPA}), (\ref{ORPA0}),
against an exact numerical solution computed by diagonalizing a Hamiltonian
in a shell-model basis.
We work in complete $0\hbar\omega$ spaces where the valence
particles are restricted to the single-particle states of a single major harmonic
oscillator shell, e.g., $1s_{1/2}$-$0d_{3/2}$-$0d_{5/2}$ for \textit{sd} or
$1p_{1/2}$-$1p_{3/2}$-$0f_{5/2}$-$0f_{7/2}$ for \textit{pf} shell.
The cores are inert, $^{16}$O for the \textit{sd} and $^{40}$Ca for the 
\textit{pf} shell. In the the valence space we 
use realistic interactions: Wildenthal ``USD'' in the \textit{sd} shell \cite{wildenthal}
and the monopole-modified Kuo-Brown ``KB3'' in the \textit{pf} shell \cite{KB3}.

As mentioned in the introduction, an operator one would like to
measure is $R^2$.  Unfortunately in a $0\hbar\omega$ shell-model
space $\langle R^2 \rangle$ is trivial. The operator $R^2$ has
two pieces, a one-body piece $r^2$, which has a constant
expectation value in a
single major harmonic oscillator shell, and a two-body piece
$r(1) \cdot r(2)$; but because $r$ is an odd-parity operator,
the two-body piece is non-zero only across major shells.
Therefore in this paper we consider other one-
and two-body operators.

\subsection{One-body operators}

Using the quasiboson approximation, Rowe derived the 
RPA one-particle densities for spherical nuclei \cite{Rowe70,ro68}; 
related formulas have been used to compute the isotope shift 
in calcium \cite{pa73}, ignoring the two-body contribution. 
Let $\rho_M = \langle c^\dagger_M c_M \rangle$ be the 
number of particles excited above the Fermi surface into the 
particle state $M$. ($M$ is not a magnetic quantum number.) Here 
$\tilde{A}_{mi,nj} = \delta_{ij} \delta_{mM} \delta_{nM}$ 
and from (\ref{ORPA0})
\begin{equation}
\rho_M = \sum_{\lambda} \sum_i |Y_{Mi,\lambda}|^2,
\end{equation}
which is Rowe's result for \textit{spherical} Hartree-Fock states. 
If the HF state is 
deformed, however, one has to take into account 
zero modes. The second term in (\ref{ORPA0}) can be simplified if 
one sums over all particle states $M$, that is, the total  
number of particles excited above the Fermi surface; in 
that case one can use Eq. (\ref{qpnorm}) and find 
\begin{equation}
\sum_M \rho_M = \sum_{\lambda} \sum_{Mi} |Y_{Mi,\lambda}|^2 - \frac{1}{2}N_{\rm zero}
\end{equation} 
where $N_{\rm zero}$ is the number of zero modes, $N_{\rm zero}=2$ for an
axisymmetric Hartree-Fock state, such as for $^{20}$Ne, and
=3 for a triaxial state, found for $^{24}$Mg. We find often leads 
to negative occupation numbers!  Note, however, 
that $\sum \rho_M$ is not an angular momentum scalar, nor even a 
spherical tensor of fixed rank, for deformed states.

Instead we considered the only scalar one-body operators in our system, 
the occupation numbers of a $j$-shell.  
Table \ref{numocc} tabulates the Hartree-Fock, RPA, and exact 
shell-model (SM) values of $n(d_{3/2})+n(s_{1/2})$. The results are 
rather poor, except for three cases with spherical Hartree-Fock states, 
$^{22}$O, $^{24}$O, and $^{48}$Ca (for which we instead 
tabulate $n(f_{7/2})$). $^{21}$O is weakly deformed and also shows a 
reasonable, albeit imperfect, improvement as one goes from 
the Hartree-Fock value to the RPA value.

\begin{table}
\caption{The number of particles in 
the $d_{3/2}+ s_{1/2}$ orbits 
($^\dagger = f_{7/2}$ orbit for $^{48}$Ca). 
$^*$ denotes nuclides with spherical 
Hartree-Fock states.}
\begin{ruledtabular}
\begin{tabular}{cccc}
  Nucleus   & HF   &  RPA &  SM \\
\hline
$^{20}$Ne & 1.60 & 1.75 & 1.60 \\
$^{22}$Ne & 1.49 & 1.64 & 1.47 \\
$^{24}$Mg & 2.13 & 1.85 & 2.31 \\
$^{28}$Si & 3.87 & 3.77 & 3.15 \\
\hline
$^{20}$O  & 0.24 & 0.46 & 0.54 \\
$^{21}$O  & 0.06 & 0.53 & 0.38 \\
$^{22}$O$^*$  & 0.00 & 0.67 & 0.54 \\
$^{24}$O$^*$  & 2.00 & 2.50 & 2.25 \\
$^{48}$Ca$^{*\dagger}$ & 8.00 & 7.73  & 7.78 \\
\hline
$^{22}$Na & 1.51 & 1.69 & 1.52 \\
$^{26}$Al & 2.29 & 2.51 & 2.18 \\
\hline
$^{19}$F  & 1.02 & 1.10 & 1.52 \\
$^{21}$F  & 0.84 & 0.56 & 0.87 \\
$^{25}$Mg & 2.21 & 3.09 & 1.96
\end{tabular}
\end{ruledtabular}
\label{numocc}
\end{table}

\begin{figure}
\centering
\includegraphics*[scale=1.0]{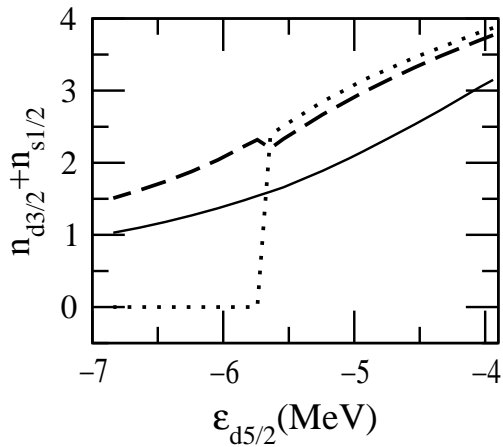}
\caption{$n(d_{3/2})+n(s_{1/2})$ in  $^{28}$Si as the $d_{5/2}$ single-particle
energy is lowered relative to the other single-particle energies.
The solid line is the (exact) shell-model value, the 
dotted line the HF value, and the dashed line the RPA value.
}
\label{numoccfig}
\end{figure}

To explore this issue further, we took $^{28}$Si and lowered 
the $d_{5/2}$ single-particle energy; eventually the 
Hartree-Fock state changes from deformed to spherical.
The results are plotted in Figure \ref{numoccfig}. 
Again we see reasonable agreement for the spherical 
region, but poor agreement in the deformed regime.

To summarize our results for one-body operators: we regain, 
for spherical Hartree-Fock states, Rowe's one-particle occupation 
numbers and get improved values over the Hartree-Fock 
occupation numbers. For deformed nuclides, however, 
the RPA value is generally worse than the HF value. 
The fault does 
not appear to lie in the corrections due to zero modes; in the next 
section, we will find that the RPA expectation value of 
$J^2$ is more accurate 
in the deformed regime than in the spherical regime.

\subsection{Two-body operators}

We now turn to two-body operators, or more properly
operators with both one- and two-body pieces.
(We investigated the pure two-body pieces but found
no qualitative differences; the pure two-body pieces
performed neither better nor worse on the whole than
the one-body pieces, which here are linear combinations
of number operators.)

In Table \ref{LS} we show results for
$S^2$ (total spin) and $L^2$ (total orbital angular momentum).
The RPA expectation value is generally a significant improvement
over the Hartree-Fock value, relative to the exact result.
On the other hand, the RPA values, while closer to the mark,  
are not in very good agreement with the exact shell-model values, 
and sometimes overcorrect to negative, nonphysical expectation values 
(this can happen because
RPA does not respect the Pauli exclusion principle).

In Table \ref{Pair} we consider the expectation value of
the pairing interaction, $P^\dagger P$ where
$P^\dagger = \sum_{j,m>0} a^\dagger_{j,m} a^\dagger_{j, -m}$, and of
$Q^2$. We also show the ratio of correlation energies
\cite{stetcu2002} which is a measure of how well the RPA binding
energy tracks the exact binding energy.  There appears to be
no correspondence: a good RPA value for the binding energy
does not correspond to a good RPA expectation value.
In particular, note the single-species (oxygen) results,
where the RPA binding energy is particularly bad; yet for
these nuclides $\langle P^\dagger P \rangle$ and
$\langle Q^2 \rangle$ are very good.

\begin{table}[h]
\begin{ruledtabular}
\caption{Ground-state expectation values of $S^2$ and $L^2$ for
several nuclei in \textit{sd} and \textit{pf} shells. For each
observable we show the SM, HF and RPA estimates. The nuclides
have been grouped into even-even, single-species, odd-odd,
and odd-A.} \label{LS}
\begin{tabular}{cccccccc}
  Nucleus &\multicolumn{3}{c}{$S^2$}& &\multicolumn{3}{c}  {$L^2$} \\
\cline{2-4} \cline{6-8}&
   HF &  RPA  & SM  & &  HF & RPA & SM  \\
\hline
 $^{20}$Ne& 0.35& 0.33 &0.26&  &15.90 &-0.25 & 0.26\\
 $^{22}$Ne&  1.48& 0.48 &0.88& &16.76 & 0.31 & 0.88\\
 $^{24}$Mg&  1.39& 1.38 &1.03& &20.65 &-1.17 & 1.03\\
 $^{26}$Mg& 2.04& 1.14 &1.45&  &18.94 &-0.34 & 1.45\\
 $^{28}$Si&  1.62& 1.28 & 1.45& &21.50 &-0.75& 1.45 \\
 $^{44}$Ti&  1.03  & 0.75  &0.64  &  &30.34 &-2.48& 0.64\\
 $^{46}$Ti&   2.24 & 1.20  & 1.36  &  & 29.72 &-2.94 & 1.36\\
 $^{48}$Cr&   3.12 & 0.99  &1.70  &  & 29.77 & 5.38 & 1.70 \\
\hline
 $^{20}$O &  1.50  & 0.45 &0.75  &  & 6.80 &  0.92& 0.75\\
 $^{22}$O &  2.40  &  -0.15&1.26  &  & 2.40 &6.36& 1.26\\
 $^{24}$O &  2.40  &  -0.27&1.29  &  & 2.39 &6.06 & 1.29\\
\hline
 $^{20}$F &  2.00 & 1.44&1.74& &14.21 & 8.09 & 3.55\\
 $^{22}$Na&  2.20& 1.90 &2.14& &21.32 & 9.08 & 8.07\\
 $^{26}$Al&  3.14& 1.96 &1.45& &29.56 &20.14& 1.45 \\
 $^{46}$V &   2.51 & 1.50  &1.36  & & 35.39 &16.33 & 1.36 \\
\hline
 $^{19}$F &  1.09& 0.80 &0.87& &12.61 & 4.39 & 0.22\\
 $^{21}$F &  2.11& 0.76 &1.52& &13.31 & 5.60 & 6.41\\
 $^{21}$Ne& 1.11& 0.44 &1.00&  &17.55 &10.11 & 3.22\\
 $^{23}$Na&  2.02& 0.88 &1.15& &18.81 & 7.46 & 3.93\\
 $^{25}$Mg&  2.04& 0.38 &1.73& &22.56 &11.77 & 7.68\\
\end{tabular}
\end{ruledtabular}
\end{table}

\begin{table}[h]
\caption{Ground-state expectation values of $P^\dagger P$ (pairing) and
$Q\cdot Q$. The final column is the ratio of the RPA correlation
energy to the shell-model correlation energy, and =1 when the
RPA binding energy is equal to the exact binding energy.} \label{Pair}
\begin{ruledtabular}
\begin{tabular}{cccccccc}
Nucleus &\multicolumn{3}{c}{Pairing}&
\multicolumn{3}{c}{$Q\cdot Q$}& ${E_{HF}-E_{RPA}}\over{E_{HF}-E_{SM}}$ \\
 \cline{2-4} \cline{5-7}
 &   HF   &  RPA  & SM    & HF & RPA &SM & \\
\hline
 $^{20}$Ne &        2.99 &    5.47  & 6.81 &  715 & 825 & 793 &0.75  \\
 $^{22}$Ne &      3.99 &    7.25  &   9.31 &876 & 1007&  944 & 0.97  \\
 $^{24}$Mg &       5.99 &   10.14  & 11.72 &  1167& 1263& 1268&0.92  \\
 $^{26}$Mg &     6.99 &   11.51  &   14.56 &  1001& 1104& 1048&0.94  \\
 $^{28}$Si &     8.99 &   12.73  &   15.16 & 1304& 1389& 1214&0.90  \\
 $^{20}$O  &       2.00 &    5.18  &  7.25 & 257 & 353 & 339 & 1.09  \\
 $^{22}$O  &       3.00 &    5.83  & 6.20 &   163 & 277 & 270 &1.67  \\
 $^{24}$O  &        4.00 &    6.52  & 6.58 & 122 & 194 & 191 & 1.83
\end{tabular}
\end{ruledtabular}
\end{table}

\begin{figure}
\centering
\includegraphics*[scale=1.0]{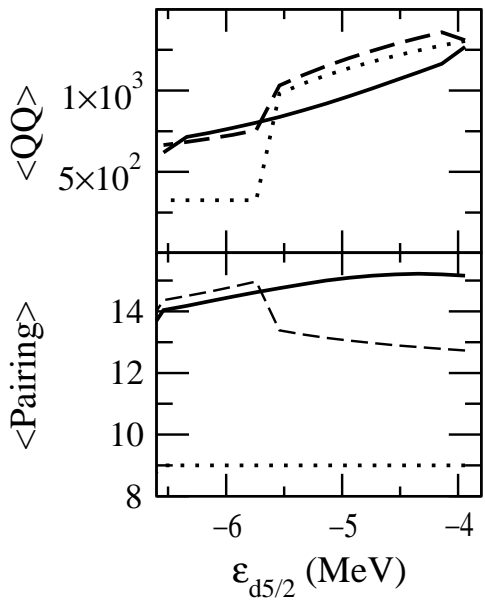}
\caption{$\langle Q^2 \rangle $, $\langle P^\dagger P \rangle $ 
in  $^{28}$Si as the $d_{5/2}$ single-particle
energy is lowered relative to the other single-particle energies.
The solid line is the (exact) shell-model value, the 
dotted line the HF value, and the dashed line the RPA value.
}
\label{Si28fig}
\end{figure}

Again we look at the transition from deformed to spherical in $^{28}$Si 
for $\langle Q^2 \rangle $ and $\langle P^\dagger P \rangle $, 
in Fig.~\ref{Si28fig}, which clearly shows the RPA values are 
in better agreement in the spherical regime than in the deformed 
regime. (As it happens, of the nuclides we investigated 
$^{28}$Si, while convenient for comparing spherical vs deformed 
regimes, is the only nuclide for which the RPA value of $Q^2$ is worse 
than the HF value, using the original Wildenthal single-particle 
energies.) This is not universal behavior; as seen in Table \ref{LS}  and 
will be seen in the next section for $J^2$, the RPA expectation 
value for some operators is better in the deformed regime.

\section{Restoration of broken symmetries?}

The random phase approximation respects broken symmetries by 
separating out exactly, as zero modes, spurious motion.  
This is sometimes interpreted as an ``approximate restoration 
of the symmetry'' \cite{ring}.  The restoration cannot 
be exact, because the RPA wavefunction is valid only in 
the vicinity of the Hartree-Fock state \cite{weneser}
and cannot be 
extrapolated to, for example, large rotation angles. 

Still, we now have a tool to further explore symmetry 
restoration, by computing Casimir operators of symmetry groups. 
Specificially, we consider $\langle J^2 \rangle$. 
Ideally, if the RPA restores a broken symmetry, one might imagine that
one either regains the exact ground state value of $\langle J^2 \rangle$
or gets very close to it.

We present our results in Table \ref{J2}.  The pattern is the same
as with other operators: $\langle J^2 \rangle$ is generally better in RPA
than in Hartree-Fock but not always very close to the exact shell-model value.
Even worse are the cases with a closed shell in HF, such 
as $^{22,24}$O: the HF value is correct, while the RPA value is
terrible!

\begin{table}[h]
\begin{ruledtabular}
\caption{Ground-state expectation value  $\langle
{J}^2\rangle$ for selected nuclides, grouped into even-even,
single-species, odd-odd, and odd-A. \label{J2}}
\begin{tabular}{cccc}
  Nucleus   & HF   &  RPA &  SM \\
\hline
   $^{20}$Ne &  16.06 &-0.45& 0\\

   $^{22}$Ne&  17.17 &-1.16& 0 \\
   $^{24}$Mg &  20.13 &-2.52& 0\\
   $^{26}$Mg &  18.61 &-1.72& 0\\
   $^{28}$Si &  20.89 &-1.99 & 0\\
$^{44}$Ti &31.65 & -3.10  &   0 \\
$^{46}$Ti &31.53 & -5.00  &   0 \\
$^{48}$Cr &29.37 &  4.72  &   0 \\
   $^{20}$O &  6.07 & 1.76 & 0 \\
   $^{22}$O &  0.00 & 7.99 & 0 \\
   $^{24}$O &  0.00 & 7.38& 0 \\
\hline
   $^{20}$F & 18.46 &12.41 & 6\\
   $^{22}$Na &  25.57 &14.57& 12\\
   $^{26}$Al &  35.98 &27.92& 0\\
$^{46}$V  &39.56 & 20.00  &   0 \\
\hline
   $^{19}$F & 15.12 & 5.52 & 0.75 \\
   $^{21}$F & 15.51 & 9.47 & 8.75  \\
   $^{21}$Ne &  19.05 &12.68& 3.75\\
   $^{23}$Na&  19.42 &11.87& 3.75 \\
   $^{25}$Mg &  23.87 &14.51& 8.75\\
\end{tabular}
\end{ruledtabular}
\end{table}

\begin{figure}
\centering
\includegraphics*[scale=1.0]{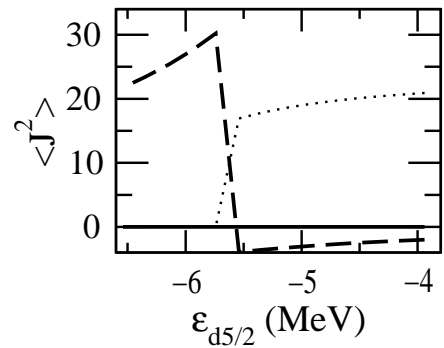}
\caption{$\langle J^2 \rangle $ in  $^{28}$Si as the $d_{5/2}$ single-particle
energy is lowered relative to the other single-particle energies.
The solid line is the (exact) shell-model value, the 
dotted line the HF value, and the dashed line the RPA value.
}
\label{J2fig}
\end{figure}

To examine this issue more closely, in Fig. \ref{J2fig} we again plot, for $^{28}$Si,
$\langle J^2 \rangle$
versus the $d_{5/2}$ single-particle energy through the transition from
deformed to spherical HF state.  The results are better for the deformed 
HF state, although we obtain slightly negative, and thus nonphysical, 
values of $\langle J^2 \rangle$. 

An additional test of symmetry restoration would be 
computation of the expectation value of a nonscalar observable, 
such as the magnetic dipole moment or electric quadrupole moment, 
for a deformed nucleus with a $J=0$ shell-model ground state. 
We have preliminary, unpublished calculations 
which suggest that indeed the RPA ground state of even-even 
nuclides retains a significant quadrupole moment, another 
piece of evidence that symmetry is incompletely restored.

There are other observables one would like to compute relevant to broken
symmetries. Aside from the Casimir itself, the dispersion 
of a Casimir would be a useful measure. For example, consider quasi-particle
RPA (QRPA), where particle number is broken in the Hartree-Fock-Bogoliubov 
state. One would like to see the QRPA value of the 
dispersion $\langle \hat{N}^2 \rangle - \langle \hat{N} \rangle^2$
move close to zero.  Another example would be proton-neutron RPA
(pnRPA) or pnQRPA, allowing protons and neutrons to mix, so that
$T_z = Z-N$ is no longer an good quantum number; this is applicable 
to beta decay. Here one might consider $\langle
\hat{T}_z^2 \rangle -\langle \hat{T}_z \rangle^2$.  Unfortunately, 
we suspect that the dispersion would also signal incomplete 
symmetry restoration.

\section{Summary and conclusion}

We derived a expression for the ground-state expectation value of
observables in the matrix formulation of RPA, and tested it against
exact shell-model calculations for selected scalar operators.
The RPA value was in general an improvement over the Hartree-Fock
value, but failed to be a consistent and reliable estimate of the exact
expectation value. Nonetheless this work should be considered a
starting point for any modified RPA calculations, such as
renormalized RPA, etc.

In particular we considered the expectation value of $J^2$. 
If one starts with a deformed Hartree-Fock state, which breaks 
rotational invariance, the RPA 
approximately restores rotational symmetry, as evinced by 
better values of $\langle J^2 \rangle.$  The results are not 
wholly satisfactory, however, as $\langle J^2 \rangle$ can 
take on unphysical (negative) values; furthermore, if one starts 
from a Hartree-Fock state with good symmetry, the HF value of $\langle J^2 \rangle$ 
is correct while the RPA value 
is large and positive, a disappointing result. 
Thus, while the RPA \textit{respects} or identifies broken symmetries
exactly, one can only characterize the \textit{restoration} of 
symmetry in the RPA as approximate and 
somewhat unreliable.  

\begin{acknowledgments}
The U.S.~Department of Energy supported this investigation through
grant DE-FG02-96ER40985.
\end{acknowledgments}

\end{document}